\documentclass[]{rstransa}
\usepackage[utf8]{inputenc}	

\usepackage[dvipsnames]{xcolor}

\usepackage{lmodern} 
\usepackage{bm}
\usepackage{caption}	
\makeatletter
\renewcommand\p@subsection{\thesection} 
\makeatother

\makeatletter
\def\itemhook{}

\def\enumhook{}

\def\deschook{}
\def\enumerate{%
  \ifnum \@enumdepth >\thr@@\@toodeep\else
    \advance\@enumdepth\@ne
    \edef\@enumctr{enum\romannumeral\the\@enumdepth}%
      \expandafter
      \list
        \csname label\@enumctr\endcsname
        {%
          \enumhook \csname enumhook\romannumeral\the\@enumdepth\endcsname%
          \usecounter\@enumctr\def\makelabel##1{\hss\llap{##1}}%
        }%
  \fi}
\def\itemize{%
  \ifnum \@itemdepth >\thr@@\@toodeep\else
    \advance\@itemdepth\@ne
    \edef\@itemitem{labelitem\romannumeral\the\@itemdepth}%
    \expandafter
    \list
      \csname\@itemitem\endcsname
      {%
        \itemhook \csname itemhook\romannumeral\the\@itemdepth\endcsname%
        \def\makelabel##1{\hss\llap{##1}}%
      }%
  \fi}

\makeatother

\renewcommand{\theenumi}{\arabic{enumi}}        

\renewcommand{\theenumii}{\alph{enumii}}

\renewcommand{\enumhook}{
  \setlength{\topsep}{0ex}
  \setlength{\itemsep}{0ex}
  \setlength{\leftmargin}{2em}
}

\renewcommand{\itemhook}{
  \setlength{\topsep}{0ex}
  \setlength{\itemsep}{0ex}
  \setlength{\leftmargin}{2em}
}

\newcommand{\im}{i}				

\newcommand{\bgamma}{\bm{\gamma}}

\newcommand{\bT}{\bm{T}}

\newcommand{\Cl}{{\rm Cl}}
\newcommand{\EE}{E}

\newcommand{\SO}{{\rm SO}}
\newcommand{\Sp}{{\rm Sp}}
\newcommand{\Spin}{{\rm Spin}}
\newcommand{\SU}{{\rm SU}}
\newcommand{\U}{{\rm U}}

\newcommand{\unit}[1]{\, {\rm #1}}		

\newcommand{\pionfig}{
    \begin{figure}[h!]
    \begin{center}
    \leavevmode
    \includegraphics[scale=.7]{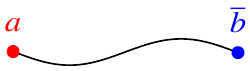}
    \caption[quark-antiquark pair joined by a string]{
    \label{pion}
Quark-antiquark pair joined by a string.
The fermions are not themselves strings,
but rather they are the endpoints of strings.
    }
    \end{center}
    \end{figure}
}

\newcommand{\stringfig}{
    \begin{figure}[t!]
    \begin{center}
    \leavevmode
    \includegraphics[scale=.7]{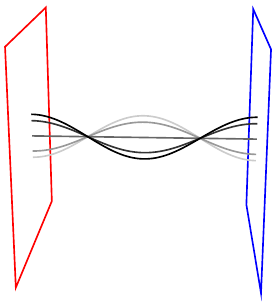}
    \hspace{1em}
    \includegraphics[scale=.7]{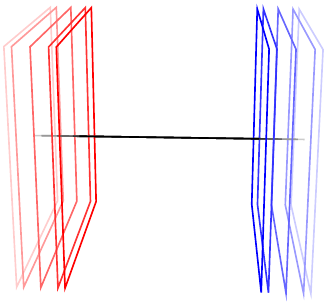}
    \hspace{1em}
    \includegraphics[scale=.7]{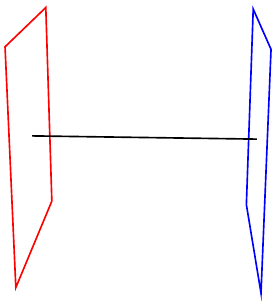}
    \caption[Oscillations of a string and its D-brane boundary]{
    \label{string}
Left: Free gauge bosons are described by transverse oscillations of the open string that leave the D-brane unexcited.
Middle: Free fermions are described by transverse oscillations of the D-brane that leave the open string unexcited.
Right: The open-string tachyon is the eigenstate of no excitation of either string or D-brane.
    }
    \end{center}
    \end{figure}
}

\newcommand{\higgsfig}{
    \begin{figure}[t!]
    \begin{center}
    \includegraphics[scale=.5]{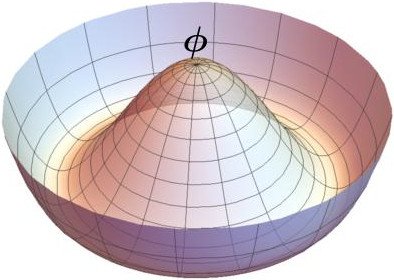}
    \caption[Vacuum Higgs field atop a Mexican-hat potential]{
    \label{higgs}
Vacuum Higgs field $\phi = 0$ sitting at the top of a Mexican-hat potential. 
    }
    \end{center}
    \end{figure}
}

\newcommand{\couplingfig}{
    \begin{figure}[t!]
    \begin{center}
    \includegraphics[scale=.625]{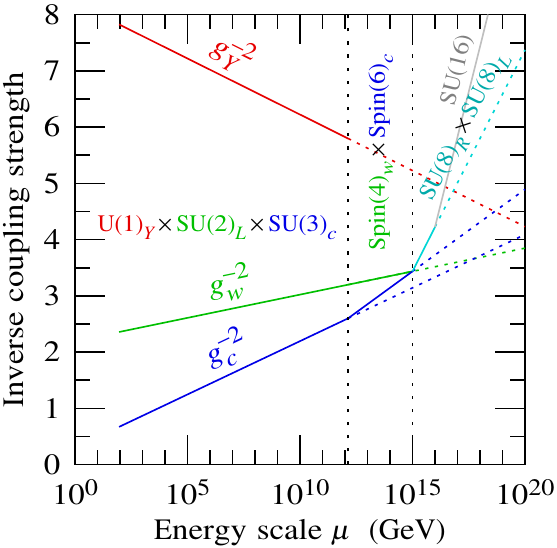}
    \includegraphics[scale=.625]{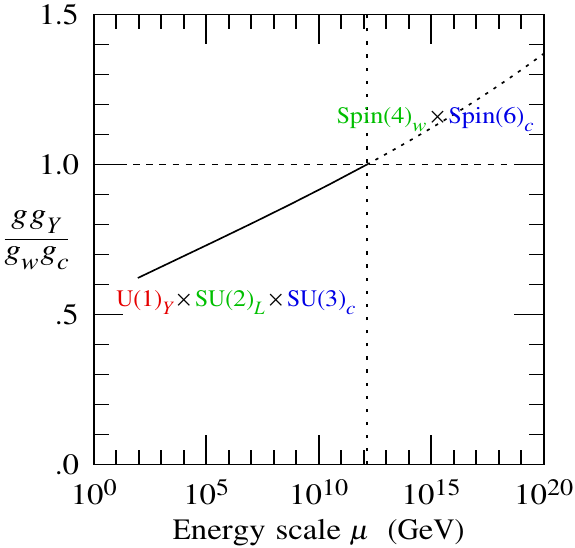}
    \caption[Running of coupling parameters with energy]{
    \label{coupling16}
(Left)
Running of the standard-model coupling parameters $g_Y$, $g_w$, and $g_c$
with renormalization energy scale $\mu$.
The solid blue line $g_c^{-2}$ tracks the standard-model
colour force $\SU(3)_c$ below the transition to the Pati-Salam group
$\Spin(4)_w \times \Spin(6)_c$ at $\mu \approx 10^{12} \unit{GeV}$,
then $\Spin(6)_c$ above the transition energy.
The solid green line $g_w^{-2}$ tracks the weak force $\SU(2)_L$
below the transition energy,
then
$\Spin(4)_w \cong \SU(2)_R \times \SU(2)_L$
above the Pati-Salam energy.
The weak and colour couplings $g_w$ and $g_c$ merge
at $\mu \approx 10^{15} \unit{GeV}$.
The cyan line tracks the gauge group $\SU(8)_R \times \SU(8)_L$
above the weak-colour merger scale $\mu = 10^{15} \unit{GeV}$.
Grand symmetry breaking from $\SU(16)$, mediated by the grand Higgs field $\bT$,
equation~(\ref{Higgsnuvac}),
could occur at any energy at or above the merger scale
$\mu = 10^{15} \unit{GeV}$;
the grey line tracks $\SU(16)$ with grand symmetry breaking taken,
arbitrarily, at
$\mu = 10^{16} \unit{GeV}$.
The lines are straight lines when $g^{-2}$
is plotted versus log energy.
(Right)
Unification of the standard model to
the Pati-Salam group $\Spin(4)_w \times \Spin(6)_c$
occurs where $g g_Y / ( g_w g_c ) = 1$
with $g \equiv \sqrt{g_w^2 + g_c^2}$,
which happens at energy scale $\mu \approx 10^{12} \unit{GeV}$
\cite[eq.~(86)]{Hamilton:2022b}.
    }
    \end{center}
    \end{figure}
}

\newcommand{\calabiyaufig}{
    \begin{figure}[t!]
    \begin{center}
    \includegraphics[scale=.35]{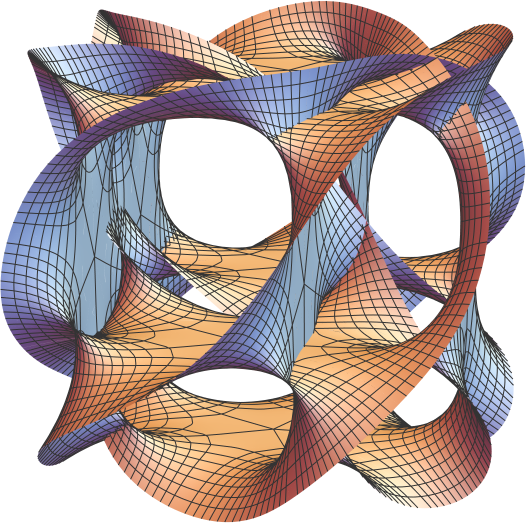}
    \end{center}
    \caption[Calabi-Yau three-fold]{
    \label{calabiyau}
The simplest class of nontrivial Calabi-Yau $n$-fold is defined by
the vanishing of a homogeneous complex polynomial of degree $n{+}2$
in a complex projective space $\mathbb{CP}^{n+1}$.
The surface shown is the simplest example of
a Calabi-Yau three-fold, $n = 3$,
a 3-dimensional cut through the 4-dimensional quintic
Fermat surface $z_1^{n+2} + z_2^{n+2} = 1$.
This image was made using Mathematica code created by Andrew Hanson
\url{https://demonstrations.wolfram.com/CalabiYauSpace/},
described in \cite{Hanson:1994}.
    }
    \end{figure}
}

\newcommand{\grouptab}{
    \begin{table*}[t!]
    \caption[The $G^2$ group and the dimensions of its representations]{
The $G^2$ group and the dimensions of its spinor ($\hat{\tfrac{1}{2}}$)
and adjoint ($\hat{2}$) representations
    }
    \begin{center}
    \label{group}
    \begin{tabular}{c@{\ \ }c@{\ \ }c@{\ \ }c@{\ \ }c}
\hline
\hline
\rule{0pt}{2.8ex}
$N$ & 
$G^2(N)$ & rank & $\hat{\tfrac{1}{2}}$ & $\hat{2}$ 
\rule[-1.2ex]{0pt}{0pt}
\\
\hline
\phantom{1}2 &
$\U(1)$ & 1 & $2 \times 1$ & $1 \times 1$  %
\\
\phantom{1}3 &
$\Sp(1)$ & 1 & $2$ & $1 \times 3$ 
\\
\phantom{1}4 &
$\Sp(1)^2$ & 2 & $2 \times 2$ & $2 \times 3$ 
\\
\phantom{1}5 &
$\Sp(2)$ & 2 & $4$ & $2 \times 5$ 
\\
\phantom{1}6 &
$\U(4)$ & 4 & $2 \times 4$ & $4 \times 4$ 
\\
\phantom{1}7 &
$\Spin(8)$ & 4 & $8$ & $4 \times 7$ 
\\
\phantom{1}8 &
$\Spin(8)^2$ & 8 & $2 \times 8$ & $8 \times 7$ 
\\
\phantom{1}9 &
$\Spin(16)$ & 8 & 16 & $8 \times 15$ 
\\
10 &
$\U(16)$ & 16 & $2 \times 16$ & $16 \times 16$ 
\\
11 &
$\Sp(16)$ & 16 & 32 & $16 \times 33$ 
\\
12 &
$\Sp(16)^2$ & 32 & $2 \times 32$ & $32 \times 33$ 
\\
13 &
$\Sp(32)$ & 32 & $64$ & $32 \times 65$ 
\\
\hline
\hline
    \end{tabular}
    \end{center}
    \end{table*}
}

\titlehead{Research}

\begin{document}

\title{A Clifford algebra approach to bosonic strings with fermionic boundaries}

\author{Andrew J. S. Hamilton$^{1}$ and Tyler McMaken$^{2}$}

\address{$^{1}$JILA and Dept.\ Astrophysical \& Planetary Sciences,
Box 440, U. Colorado Boulder, CO 80309, USA\\
$^{2}$University of Mary, 7500 University Drive, Bismarck, ND 58504, USA}

\subject{string theory, algebra, high-energy physics, mathematical physics}


\corres{Andrew J. S. Hamilton\\
\email{Andrew.Hamilton@colorado.edu}}

\begin{abstract}
The Brauer-Weyl (1935) theorem states that the algebra of tensor products of
spinors is isomorphic to the Clifford (geometric) algebra,
in any number of spacetime dimensions.
The algebraic relation between fermions and bosons seen in the standard
model agrees with the Brauer-Weyl theorem, but not with supersymmetry.
It is commonly argued that string theory requires supersymmetry,
on the grounds that nonsupersymmetric bosonic string theory
(1) does not admit fermions,
and (2) has an unstable tachyonic ground state.
This paper rebuts those standard objections,
pointing out that bosonic string theory admits ferm\-ions
lying on a D-brane boundary of open strings,
and that the open-string tachyon has all the properties of a Higgs field.
The original 1970s formulation of bosonic string theory,
re-envisaged as a theory of all the forces, not just the strong force,
fits the standard model beautifully.
\end{abstract}


\begin{fmtext}
\section{Introduction}
\label{intro-sec}

We are fans of string theory.
To us, and to many others,
string theory seems to be the most promising theory of quantum gravity.
Whereas the standard quantum field theory of particles
famously does not admit a renormalizable quantum theory of gravity,
the simple trick of allowing fields to be strings instead of particles
automatically yields a potentially ultraviolet-complete quantum theory,
with spin~2 gravitons encoded in the vibrations of closed strings.
And string theory naturally admits the other forces of nature,
with spin~1 gauge bosons encoded in the vibrations of open strings.

\end{fmtext}

\maketitle

The textbooks
\cite{Green:1987a,Polchinski:1998a,Becker:2007}
assured us that supersymmetry was essential to string theory.
There were compelling arguments that
supersymmetric partners would be found
at the Large Hadron Collider not far above a TeV
\cite{Martin:1998}.
Those predictions were not borne out
\cite{Boubaa:2020,Bose:2022,Fox:2022,Constantin:2025}.
String theory has been left in an uncertain state.

The first goal of this paper is to refute the textbook
claim that supersymmetry is essential to string theory.
Supersymmetry posits a certain algebraic relation between
fermions and bosons
\cite{Sohnius:1985}.
The problem is that this relation disagrees with the
actual algebraic relation between fermions and bosons
seen in the standard model of physics.
In the standard model,
fermions fill a spinor representation of the gauge group,
while bosons fill the adjoint representation of the group.

The relation between fermions and bosons seen in the standard model
is encapsulated in the Brauer-Weyl (1935) theorem \cite{Brauer:1935},
which states that the algebra of outer (tensor) products of
spinors in $K{+}M$ spacetime dimensions is isomorphic to the Clifford algebra
$\Cl(K,M)$,
in any number of spacetime dimensions
\cite{Hamilton:2023a}.
%
The tension between supersymmetry and the Brauer-Weyl theorem
has bothered the authors for many years.
It has puzzled us that string theorists
did not highlight the tension.
In the end, we decided that the only way forward was to bite the bullet,
learn string theory,
and figure out for ourselves whether supersymmetry is essential to string theory.
Our conclusion: it is not.

The alternative to supersymmetric string theory is bosonic string theory,
which is expounded thoroughly in all the textbooks.
The two principal arguments that bosonic string theory is inadmissible are:
\begin{enumerate}
\item
Only supersymmetric strings admit fermions;
\item
The ground state of the bosonic string is tachyonic (negative mass-squared).
\end{enumerate}
Subsection~\ref{admitfermion-sec} rebuts the first argument,
while subsection~\ref{opentachyon-sec} argues that the open-string tachyon
is simply the Higgs field.

Section~\ref{chanpaton-sec}
shows that the Chan-Paton condition on the gauge group of a bosonic string
theory requires the internal gauge group (before any symmetry breaking)
to be $\SU(16)$, Table~\ref{group}.
Bosonic string theory requires 26 spacetime dimensions,
and \S\ref{26dim-sec} shows that the proposed Clifford(11,1) string theory
can satisfy that condition without contrivance.
Section~\ref{calabiyau-sec} shows that
the 10~dimensional compact fermionic D-brane must be the product of a torus
and a Calabi-Yau 3-fold at the present time.
Section~\ref{conclusion-sec} summarizes briefly.

\section{No, supersymmetry is not essential to string theory}

\subsection{Yes, bosonic string theory does admit fermions}
\label{admitfermion-sec}

\pionfig

The textbook assertion that only supersymmetric strings admit fermions
requires careful parsing.
The precise statement (echoed in all the texts) is
``the spectrum (of both open and closed strings) does not contain fermions''
\cite[Ch.~4, p.~109]{Becker:2007}.
What that means is that {\em if\/} fermions are, like bosons,
taken to be excitations of strings,
then those strings must exhibit supersymmetry.
However,
the earliest (1970) version \cite{Susskind:1970} of string theory
envisaged fermions not as strings themselves,
but rather as the endpoints of strings.
Specifically, the strong force was conjectured to be mediated by mesons (pions)
conceived as quark-antiquark pairs joined by a string,
as illustrated in Figure~\ref{pion}.
The theory was abandoned in favour of quantum chromodynamics (QCD) in the
mid 1970s,
because string theory predicted hadron scattering amplitudes
at large transverse momenta that declined exponentially 
at energies above the string (strong, QCD) scale,
whereas QCD correctly reproduced the observed milder power-law decay
\cite{Miller:1972}.

The nature of the endpoints of open strings remained obscure until 1995,
when Polchinski \cite{Polchinski:1995mt,Polchinski:1996}
discovered D-branes,
which are dynamical surfaces upon which open strings can end.
D-branes exist in both bosonic string theory and superstring theory
\cite[\S8.7]{Polchinski:1998a}, \cite{Sen:1999a}.
The year 1995 was a decade after the ``first superstring revolution,''
and it coincided with the ``second superstring revolution.''
The first superstring revolution was
precipitated by Green \& Schwarz's (1984) \cite{Green:1984} discovery
of a remarkable cancellation of quantum anomalies in
Type~I superstring theory with an $\SO(32)$ gauge group.
The second superstring revolution was brought about by the realization
\cite{Witten:1995ex}
that all the five known 10~dimensional supersymmetric string theories,
along with 11~dimensional supergravity,
were related by a web of dualities,
connected through a mythic 11~dimensional theory Witten dubbed ``M-theory''.


\stringfig

In bosonic open string theory with fermionic endpoints,
free gauge bosons are eigenstates of transverse oscillations of the string,
while free fermions are eigenstates of transverse oscillations
of the fermionic D-brane, as illustrated in Figure~\ref{string}.
Massless gauge bosons appear at the first excited level of the string,
and there is a Kaluza-Klein tower of massive eigenstates at higher levels.

\subsection{The open-string tachyon is the Higgs field}
\label{opentachyon-sec}

The ground state of the open string is tachyonic,
meaning it has negative mass-squared.
What this means is just that the potential energy curve of the tachyon
has a quadratic core with a negative sign,
opposite to that of a simple harmonic oscillator.
Historically the tachyonic ground state of the open string
was a major stumbling block.
It meant that the ground state of the theory was unstable.
It could not describe the apparently stable vacuum today.

The open-string tachyon is not nothing.
Rather, it is the eigenstate of the open string corresponding
to no oscillation at all, neither string nor its D-brane endpoints,
as illustrated in the right panel of Figure~\ref{string}.
Like other open strings,
the open-string tachyon has endpoints on the D-brane boundary,
and it transforms in the adjoint representation of the gauge group
\cite{Marcus:1987,Angelantonj:2002}, \cite[\S6.5]{Polchinski:1998a}.
But whereas massless gauge bosons carry charge and unit spin,
the open-string tachyon has zero charge and zero spin.

\higgsfig

In standard particle field theory,
the nominal vacuum state of a scalar field $\phi$,
where the scalar field vanishes, $\phi = 0$,
may be tachyonic, while the true vacuum state,
where the energy density has a (meta)stable minimum,
is one where the scalar field is displaced to a finite value
$\langle \phi \rangle$.
This is the idea behind the Higgs mechanism
\cite{Englert:1964,Higgs:1964},
which is the basis of the successful theory of electroweak symmetry breaking
\cite{Weinberg:1967},
and which is the inspiration for the successful theory of cosmological
inflation driven by a tachyonic scalar field
possibly at the grand unified scale
\cite{Guth:1981,Linde:1982,Akrami:2020inf,Ade:2022}.
In the electroweak Higgs field,
a negative mass-squared quadratic term $- m^2 \phi^2$
in the Higgs potential
is stabilized by a positive quartic term $\lambda \phi^4$,
leading to the `Mexican-hat' potential
illustrated in Figure~\ref{higgs}.
The potential has a stable minimum at
$\langle \phi \rangle = m / \sqrt{2\lambda}$.

The properties of the open-string tachyon resemble those of the
Higgs field of the standard model:
it is a multiplet of a symmetry group;
it is a Lorentz scalar;
and in its vacuum state $\phi = 0$
it sits atop a tachyonic potential,
Figure~\ref{higgs}.

It is an obvious idea that the open-string tachyon might be a (the) Higgs field.
A key question is whether the potential energy curve of the tachyon
has a stable minimum at a nonzero expectation value.
To leading order,
the tree-level potential of the tachyon multiplet $\phi$ contains
quadratic $\phi^2$ and cubic $\phi^3$ terms,
e.g.\ \cite[eq.~(6.5.16)]{Polchinski:1998a}.
\cite{Kostelecky:1988},
building on the computation of off-shell string-scattering amplitudes by
\cite{Samuel:1988},
carried out the first 4-point tree-level calculation of the quartic
$\phi^4$ term,
finding that the coefficient of the quartic was unfortunately negative,
therefore destabilizing.
The computation was generalized to one-loop by \cite{Bluhm:1989}.

The 1995 discovery of D-branes provoked renewed interest in open strings
and the fate of the open-string tachyon.
Sen
\cite{Sen:1998,Sen:1999b,Sen:2005}
influentially conjectured that tachyonic instability would lead to decay
of the D-brane boundary of the open string.
\cite{Gerasimov:2000}
re-examined the older computations of the tachyonic potential
in the light of \cite{Witten:1992}'s open string theory,
finding an exact solution for the potential that tended to $-\infty$
in all directions, implying irreparable instability.

However, there is something missing from all the calculations of
the potential energy curve of the open-string tachyon,
namely the coupling to fermions.
In all the calculations, fermion fields are set identically to zero.
\cite{Kostelecky:1988} justify the neglect of fermions on the grounds that the
``open bosonic string [...] cannot describe the world since it contains
no fermions'', a statement challenged in \S\ref{admitfermion-sec} above.

\cite[p.~10]{Sen:2000} assert explicitly that the open-string tachyon
cannot be a Higgs field:
``Since the tachyon is neutral,
the gauge field cannot acquire mass via the Higgs mechanism.''
It is true that the presence of a Higgs field does not endow the vacuum
with charge.
Nevertheless, the component of the Higgs multiplet
that acquires an expectation value carries charge,
and the Higgs field generates mass for the gauge bosons and fermions
to which that expectation value couples.
The electroweak Higgs field for example carries (broken) weak charge,
and it couples to $W^\pm$ and $Z$ weak gauge bosons,
and to all the fermions of the standard model.
The masses of gauge and fermion fields generated by a Higgs field
are necessarily positive, and proportional to the expectation value
of the Higgs field.
In particular, the mass of the most massive fermion, the $173 \unit{GeV}$ top quark,
plays a critical role in the potential energy curve of the Higgs field
\cite{Sher:1989,Degrassi:2012,Buttazzo:2013,Martin:2015,Agrawal:2020,Reyes:2022,Hiller:2024}.

In the standard model,
the decay of a Higgs field causes symmetry breaking of the gauge
group to a subgroup.
The holonomy group of a manifold is the group of transformations
of the tangent space when parallel transported around all possible piecewise
differentiable closed loops
\cite{Nomizu:1961}.
The tangent space of the fermionic D-brane transforms under the
gauge group.
The holonomy group of the fermionic D-brane must be a subgroup
of whatever symmetries remain after symmetry breaking.
Thus Higgs symmetry breaking leads not to decay of a D-brane,
but rather to a reduction in its holonomy group.

What needs to happen is that the potential energy curve of the
open-string tachyon should be recalculated including the couplings
to fermions and gauge fields.
If the potential energy curve shows a stable minimum,
then bosonic open string theory is back in business.
If not, then the proposed bosonic string theory is ruled out.

\section{The Chan-Paton condition on the gauge group}
\label{chanpaton-sec}

String theory began not as a theory of strings,
but rather with the conjecture by Veneziano \cite{Veneziano:1968}
that the scattering of hadrons was described by a certain
quantum mechanical amplitude.
The conjecture became relevant to strings when
\cite{Susskind:1970} and a few others realized that the Veneziano amplitude
described the tree-level scattering of open bosonic strings.

The cyclic symmetry of the Veneziano amplitude
imposes a tight constraint,
the Chan-Paton \cite{Paton:1969} condition,
on the relation between the group representations
of the bosonic string and of the charges at its endpoints.
The Chan-Paton condition is that,
if the endpoints of an open string transform
in a certain representation $a$ of dimension $n$ of the gauge group,
then the open string must transform in a tensor representation
of that representation.
If the string is oriented, as in the original quark-antiquark model,
Figure~\ref{pion},
then the tensor representation is $a\bar{b}$ of dimension $n^2$.
If the string is unoriented, as in Type~I superstring theory,
then the string is unchanged up to a sign by swapping its endpoints,
and the tensor representation is either antisymmetric $[ab]$,
dimension $\tfrac{1}{2} n(n{-}1)$,
or symmetric $(ab)$,
dimension $\tfrac{1}{2} n(n{+}1)$.
An exposition of the Chan-Paton condition can be found in any of the textbooks
\cite[\S6.1.1, p.~292]{Green:1987a},
\cite[\S6.5, p.~184]{Polchinski:1998a},
\cite[\S6.1, p.~196]{Becker:2007}.

The Brauer-Weyl (1935) theorem \cite{Brauer:1935} states what the
algebra of tensor products of spinors is:
it is the Clifford algebra, in any number of dimensions.
The tension in the textbooks between
the assertion that bosonic strings cannot admit fermions,
and the dangling mathematical and physical evidence
that the endpoints can be fermionic, is palpable.
In a footnote on p.~293 while discussing the Chan-Paton condition,
\cite{Green:1987a} write:
``By `quarks' and `antiquarks' [quotes per the original] we only mean
labels representing certain quantum numbers.
To attach mass or spin at the ends of open strings would ruin the
analysis described in the previous chapters.''
The ``analysis in the previous chapters''
covers the spectrum of excitations of strings,
not what lies on their endpoints.
As illustrated in Figure~\ref{string},
in bosonic string theory,
a free gauge boson,
such as a red-antiblue gluon
$\textcolor[rgb]{1,0,0}{r}\textcolor[rgb]{0,0,1}{\bar{b}}$,
is an excitation of the string that leaves its D-brane boundary unexcited,
while a free quark is an excitation of the D-brane that leaves the
string unexcited.
The gauge bosons and the fermions do interact;
but those interactions should not be confused with the free eigenstates
that comprise the in- and out-states of a Feynman diagram.


If nonsupersymmetric bosonic string theory is to accommodate the standard model,
then fermions must live on the endpoints of strings,
since the alternative, that fermions are themselves strings,
requires supersymmetry.
As has been known since 1975
\cite{Georgi:1975,Fritzsch:1975},
a generation of fermions of the standard model
fit neatly into the spinor representation of $\Spin(10)$
(the covering group of $\SO(10)$),
with $2^{[10/2]} = 32$ components
(including the unobserved right-handed neutrino,
whose predicted presence is to be considered a plus given that
the left-handed neutrino apparently has mass,
and pure-handed spinors are necessarily massless).
The Brauer-Weyl theorem states that the algebra of outer (tensor)
products of the spinor representation is isomorphic to the Clifford algebra.

More generally,
the spinor representation of $\Spin(N)$ has dimension $n = 2^{[N/2]}$
(technically, if $N$ is even, then there are two spinor representations
each of dimension $n/2$, one being the conjugate of the other).
The Chan-Paton condition requires that the gauge group associated
with $\Spin(N)$ spinors has dimension $n^2$ if the string is oriented,
or $n ( n {\pm} 1 ) / 2$ if the string is unoriented,
with $+$ for symmetric, $-$ for antisymmetric.
For $N \geq 6$,
this Chan-Paton dimension is larger than the dimension of $\Spin(N)$,
which is generated by the bivectors of the Clifford algebra $\Cl(N)$,
and has dimension $N(N{-}1)/2$.

\grouptab

Let $\hat{p}$ denote multivectors of grade $p$~(mod~4),
\begin{equation}
  \hat{p} \equiv p~(\textrm{mod}~4)
  \ .
\end{equation}
Physical excitations of strings are transverse,
and the Chan-Paton condition applies only to the transverse dimensions,
not to the time and longitudinal dimensions.
If the fermions transform in a spinor representation of a spin group $\Spin(N)$
in $N$ transverse dimensions, then the Chan-Paton condition
implies that the gauge group
is the group $G^2(N)$ \cite{Shirokov:2016}
generated by multivectors of grade $\hat{2}$.
Table~\ref{group} gives the group
$G^2(N)$ in $N$ transverse dimensions
\cite{Shirokov:2016},
and the dimensions of its (not necessarily irreducible) spinor
($\smash{\hat{\frac{1}{2}}}$) and adjoint ($\hat{2}$) representations.
In all cases, in any number of dimensions,
the spinor representation $\smash{\hat{\tfrac{1}{2}}}$
is the fundamental representation of the group,
and the adjoint representation $\hat{2}$ is a tensor representation
of the spinor representation, as required by Chan-Paton.
The notation for the groups in Table~\ref{group}
conforms to that of \cite[p.~150]{Cornwell:1997}.
The $\Spin(N)$ groups are covering groups of
special orthogonal groups $\SO(N)$.

For the $N = 10$ internal dimensions preferred by the standard model,
the gauge group from Table~\ref{group} is $G^2(10) = \U(16)$,
which contains $\Spin(10)$ as a subgroup.
The $\U(1)_{yzrgb}$ factor of $\U(16) = \U(1)_{yzrgb} \times \SU(16)$
is generated by the 10~dimensional pseudoscalar $I_{yzrgb}$ of $\Cl(10)$,
and can be interpreted as generating a transverse spatial rotation that,
in compliance with Coleman-Mandula theorem
\cite{Coleman:1967,Mandula:2015},
commutes with the remaining internal $\SU(16)$ group.

The conclusion is that
the internal gauge group consistent with the fermions of the standard model
and satisfying the Chan-Paton condition is $\SU(16)$.
Of course, this group must be thoroughly broken (by Higgs fields)
to arrive at the standard-model gauge group today.
$\SU(16)$ must be interpreted as the primordial gauge group,
before any symmetry breaking took place.

Each spinor of $\Spin(10)$ is actually a Weyl (massless, chiral) spinor,
with two real components,
so each fermion generation actually contains $2^6 = 64$ real components.
The Weyl degrees of freedom can be accommodated by enlarging the
Clifford algebra from $\Cl(10)$ to $\Cl(11,1)$
\cite{Hamilton:2022b}.
The reason it is possible to fit the 10 compact dimensions of $\Spin(10)$
and the 3+1 large dimensions of spacetime
inside an algebra of just 11+1 dimensions
is the well-known fact that $\Spin(10)$ chirality
(right- or left-handed from the perspective of weak interactions)
coincides with Dirac chirality
(right- or left-handed from the perspective of the Lorentz group),
implying a redundancy of the algebras.
The standard model algebra and the Dirac spacetime algebra
are commuting subalgebras of $\Cl(11,1)$
\cite{Hamilton:2022b},
in compliance with the Coleman-Mandula no-go theorem
\cite{Coleman:1967,Mandula:2015}.


We choose not to distract the reader with further details of the algebra,
which will appear elsewhere;
the enterprising reader can figure it out for themself.
The first order of business is to persuade the reader that
supersymmetry is not essential to string theory.
When supersymmetry is discarded,
there emerges a bosonic string theory
that fits beautifully with the standard model.

\section{A 26~dimensional theory}
\label{26dim-sec}

Superstring theory must live in 10 spacetime dimensions
\cite{Polyakov:1981b},
while bosonic string theory must live in 26 spacetime dimensions
\cite{Goddard:1973,Polyakov:1981},
because only in those dimensions do conformal anomalies vanish
\cite{Friedan:1986,Fujikawa:1988}.
The fermionic D-brane occupies 12 of the 26 dimensions.
What about the remaining $26{-}12 = 14$ dimensions?

The precedent for fermions to occupy a subset of bosonic dimensions
is heterotic superstring theory
\cite{Gross:1985a},
which combines left-moving bosonic strings in 26 dimensions
with right-moving fermionic strings in 10 dimensions
by compactifying $26{-}10 = 16$ of the 26 dimensions on a maximal torus of a
compact Lie group of rank 16
(maximal means that the dimension of the torus equals the rank of the group).
The compact Lie group is the compact gauge group of the theory,
which for heterotic theory had to be either $\Spin(32) / \mathbb{Z}_2$
or $\EE_8 \times \EE_8$.

In the present case the compact internal gauge group is $\SU(16)$,
which has rank 15.
The 12 dimensions of the fermionic D-brane plus the 15 dimensions of $\SU(16)$
make 27 dimensions,
one more than the canonical 26 required by bosonic string theory.
This is reminiscent of M-theory, which likewise has one more than
the canonical 10 dimensions required by superstring theory.

However, the primordial $\SU(16)$ gauge symmetry is thoroughly broken in
today's low-energy era.
Previously, \cite{Hamilton:2022b} examined
symmetry breaking starting from $\Spin(11,1)$,
concluding that there was a unique minimal symmetry-breaking chain
down to the standard model,
and an associated minimal Higgs multiplet transforming in the adjoint
representation of the gauge group.
Essentially the same arguments apply in the present case,
with the same fermion content, but with an enlarged gauge group.
The running of standard model coupling parameters implies
unification of the standard model to
the Pati-Salam group $\Spin(4)_w \times \Spin(6)_c$ \cite{Pati:1974}
at an energy scale $\mu \approx 10^{12} \unit{GeV}$,
followed by merging of the weak and colour coupling parameters at
$\mu \approx 10^{15} \unit{GeV}$,
as illustrated in Figure~\ref{coupling16},
which is an updated version of Figure~3 of \cite{Hamilton:2022b}.

\couplingfig

\cite{Hamilton:2022b} concluded that grand Higgs field
that causes grand symmetry breaking
by acquiring a nonzero expectation value $\langle \bT \rangle$ is
\cite[eq.~(46)]{Hamilton:2022b}
proportional to the $t$-bivector $\im \bgamma^+_t \bgamma^-_t$
multiplied by the colour chiral operator $\varkappa_{rgb}$,
\begin{equation}
\label{Higgsnuvac}
  \langle \bT \rangle
  =
  - \im
  \langle T \rangle
  \bgamma^+_t \bgamma^-_t
  \varkappa_{rgb}
  \ .
\end{equation}
The grand Higgs field~(\ref{Higgsnuvac}) is picked out by the property that
it is a Lorentz-invariant scalar field that
breaks $\Spin(10)$ to
the Pati-Salam group $\Spin(4)_w \times \Spin(6)_c$ \cite{Pati:1974}:
the grand Higgs field $\langle \bT \rangle$ commutes with all the generators of
the Pati-Salam group,
and fails to commute with the generators of $\Spin(10)$ not in
the Pati-Salam group.

The pre-grand internal gauge group considered in the present paper
is not $\Spin(10)$,
but rather the subgroup $\SU(16)$ of $G^2(10)$.
The grand Higgs field $\langle \bT \rangle$, equation~(\ref{Higgsnuvac}),
preserves the generators of $\SU(16)$ that commute with $I_{rgb}$,
and breaks the generators of $\SU(16)$ that fail to commute with $I_{rgb}$,
breaking $\SU(16)$ to
\begin{equation}
\label{SU16SU8SU8}
  \SU(16)
  \rightarrow
  \U(1)_{rgb} \times \SU(8)_R \times \SU(8)_L
  \ .
\end{equation}
where $\U(1)_{rgb}$ is the group generated by $I_{rgb}$.
The remaining compact group $\SU(8)_R \times \SU(8)_L$ has rank 14,
for a total number $12{+}14 = 26$ of dimensions,
which agrees with the number of dimensions required by bosonic string theory.

Whereas \cite{Hamilton:2022b} inferred that grand unification
to $\Spin(11,1)$ (or possibly to $\Spin(10,1)$) occurred at
$\mu \approx 10^{15} \unit{GeV}$,
in the present case merger to $\SU(8)_R \times \SU(8)_L$
is merely a precursor to unification to the primordial gauge group $\SU(16)$.
The coupling parameters of $\SU(8)_R$ and $\SU(8)_L$ have the same running,
so grand unification to $\SU(16)$ could occur at any energy above
$10^{15} \unit{GeV}$.
Figure~\ref{coupling16} illustrates grand unification to $\SU(16)$
at $10^{16} \unit{GeV}$.

Details of the algebra will appear elsewhere.
The takeaway is that a 26~dimensional theory,
formed from the 12 dimensions of the fermionic D-brane
plus 14 compact dimensions from a maximal torus of the rank 14 group
$\SU(8)_R \times \SU(8)_L$,
emerges without contrivance.
In the meantime, a couple more comments can be made.

Firstly, as explained by \cite{Gross:1985a},
string theory requires that a compactification torus be self-dual
(invariant under exchange $\tau \leftrightarrow \sigma$ of the
string coordinates),
which imposes a condition on the gauge group.
It can be shown that both $\SU(16)$ and $\SU(8)$ satisfy that condition.

Secondly, a consistent gauge group must be free of quantum
(gauge+grav\-itat\-ional) anomalies.
In $4n$ dimensions,
a sufficient condition for the fermionic contribution to gauge anomalies
to vanish is that right- and left-handed fermions transform
in matching representations of the gauge group, such that
the fermions could (in principle) be combined into massive Dirac fermions
\cite{Alvarez-Gaume:1983b,Bilal:2008,Alvarez-Gaume:2022}.
This is precisely the situation in $\SU(8)_R \times \SU(8)_L$.
Indeed it is the same underlying reason that the standard model is
free of gauge anomalies.
There is a possible mixed gauge-gravitational anomaly,
but that mixed anomaly turns out to vanish in $4n$ dimensions
for subgroups of groups satisfying the Chan-Paton condition.

\section{The fermionic D-brane is a Calabi-Yau manifold}
\label{calabiyau-sec}

If the open-string tachyon is the Higgs field,
as proposed in \S\ref{opentachyon-sec},
then decay of the tachyon leads to symmetry breaking,
as in standard Higgs theory.
The holonomy group of the D-brane must be a subgroup of the symmetry group.

Spinors have a natural complex structure, and the Clifford algebra
generated by outer products of spinors inherits that complex structure.
A complex manifold of $n$ complex dimensions
($2n$ real dimensions)
with $\U(n)$ holonomy
is called a K{\"a}hler $n$-fold,
while one
with $\SU(n)$ holonomy
is called a Calabi-Yau $n$-fold,
e.g.\ \cite{Moroianu:2007}.
Calabi-Yau manifolds
have the property that they are Ricci-flat
(carry zero energy-momentum),
which famously makes them prime suspects for the hypothesized
6 hidden dimensions of 10~dimensional string theory
\cite{Candelas:1985}.
Figure~\ref{calabiyau} illustrates the simplest example of a Calabi-Yau
3-fold.

\calabiyaufig

In the present case,
the fermionic D-brane has 10 compact dimensions.
If the D-brane is a Riemannian manifold,
then its holonomy group can be no greater than $\Spin(10)$.
The de Rham theorem \cite{deRham:1952} states that if the holonomy group
of a simply-connected Riemannian manifold is a direct product,
then the manifold must be a direct product.
Therefore,
after symmetry breaking to the Pati-Salam group $\Spin(4)_w \times \Spin(6)_c$,
the manifold becomes a direct product of a weak 2-fold with a colour 3-fold.
After symmetry breaking to the standard model
$\U(1)_Y \times \SU(2)_L \times \SU(3)_R$,
the manifold becomes a direct product of a weak Calabi-Yau 2-fold
with a colour Calabi-Yau 3-fold
(the $\U(1)_Y$ factor, which is not simply-connected,
and which operates on both weak and colour components,
makes no additional contribution to the manifold).
After further symmetry breaking to the present-day low energy group
$\U(1)_Q \times \SU(3)_c$,
the manifold becomes a direct product of a weak flat 2-fold
with a colour Calabi-Yau 3-fold
(the $\U(1)_Q$ factor again does not contribute).

If the fermionic D-brane is a Calabi-Yau manifold,
carrying zero energy-momentum,
then there is no lower energy state to which the D-brane can decay.
A vacuum containing a Calabi-Yau D-brane could potentially be stable.
It is not known whether the dark energy today is stable or unstable,
although it is evidently stable over cosmological timescales.
The first year of observation of large scale galaxy clustering
with the Dark Energy Spectroscopic Instrument (DESI) at Kitt Peak
\cite{Adame:2024}
shows a slight preference for decay of dark energy,
consistent with an earlier indication from five years
of Type 1a supernova data from the completed Dark Energy Survey (DES)
\cite{Abbott:2024}.

\section{Conclusion}
\label{conclusion-sec}

The algebra of bosonic string theory fits the standard model
and general relativity.
The algebra of supersymmetry does not.
Unburdened by supersymmetry,
bosonic string theory has the potential to bring string theory back
into the realm of testable physics accessible to present-day
observation and experiment.

\enlargethispage{20pt}


%
%
%
%
%
%
%
%
%
%
%
%



\providecommand{\noopsort}[1]{}

\end{document}